\documentclass[aps,showpacs,twocolumn,superscriptaddress]{revtex4}

\usepackage{graphicx}
\usepackage{dcolumn}

\begin{document}

\title{Faddeev-Yakubovsky search for $^{~~4}_{\Lambda\Lambda}$H} 

\author{I.N.~Filikhin} 
\affiliation{Racah Institute of Physics, The Hebrew University, 
Jerusalem 91904, Israel\vspace*{1ex}} 
\affiliation{Department of Mathematical and Computational Physics, 
St.Petersburg State University, 198504 St.Petersburg, Russia\vspace*{1ex}} 

\author{A.~Gal}
\affiliation{Racah Institute of Physics, The Hebrew University, 
Jerusalem 91904, Israel\vspace*{1ex}} 

\begin{abstract}
\rule{0ex}{3ex} 

Evidence for particle stability of $_{\Lambda\Lambda}^{~~4}$H 
has been suggested by the BNL-AGS E906 experiment. We report on 
Faddeev-Yakubovsky calculations for the four-body $\Lambda\Lambda pn$ 
system using $\Lambda N$ interactions which reproduce the observed 
binding energy of $_{\Lambda}^{3}{\rm H}({\frac{1}{2}}^+)$ within 
a Faddeev calculation for the $\Lambda pn$ subsystem. 
No $_{\Lambda\Lambda}^{~~4}$H bound state is found over a wide range 
of $\Lambda \Lambda$ interaction strengths, although the Faddeev 
equations for a three-body $\Lambda\Lambda d$ model of 
$_{\Lambda\Lambda}^{~~4}$H admit a $1^+$ bound state for as weak 
a $\Lambda \Lambda$ interaction strength as required to reproduce 
$B_{\Lambda\Lambda}(_{\Lambda\Lambda}^{~~6}\rm He)$. 
\end{abstract} 
\pacs{21.80.+a, 21.10.Dr, 21.10.Gv, 21.45.+v} 

\maketitle 

\section{Introduction and input} 

Information on hyperon-hyperon interactions is not readily available 
from experiments in free space. It is almost exclusively limited to 
the study of strangeness $S$=$-2$ hypernuclear systems, only handful 
of which have been identified to date. This information is crucial for 
extrapolating into multi-strange hadronic matter, for both finite 
systems and in bulk (Ref. \cite{SBG00} and references cited therein). 
Until recently only three candidates, identified in emulsion 
experiments \cite{Dan63,Pro66,Aok91}, existed for $\Lambda\Lambda$ 
hypernuclei. The $\Lambda\Lambda$ binding energies 
deduced from these events indicated that the $\Lambda\Lambda$ 
interaction is strongly attractive in the $^{1}S_0$ channel \cite{BUs87}, 
in fact considerably stronger than the $\Lambda N$ interaction 
deduced from single-$\Lambda$ hypernuclei, and this seemed at 
odds with the natural expectation borne out in one-boson-exchange 
(OBE) models using flavor SU(3) symmetry or within the naive 
quark model. For example, the recent Nijmegen soft-core (NSC97) 
model \cite{RSY99,SRi99} yields 
\begin{equation} 
\label{eq:V1} 
{\bar V}_{\Lambda\Lambda} \ll {\bar V}_{\Lambda N} \ll 
{\bar V}_{NN} \;\; 
\end{equation} 
for the strength $\bar V$ of these essentially attractive 
interactions. It is gratifying then that the recent unambiguous 
identification of $_{\Lambda\Lambda}^{~~6}$He in the KEK 
hybrid-emulsion experiment E373 \cite{Tak01}, yielding binding 
energy substantially lower than that deduced from the older 
dubious event \cite{Pro66}, is consistent with 
a scattering length $a_{\Lambda\Lambda}$$\sim$$-0.5$ fm 
\cite{FGa02}, indicating a considerably weaker $\Lambda\Lambda$ 
interaction than that specified by $a_{\Lambda N}$$\sim$$-2$ fm 
\cite{RSY99} for the $\Lambda N$ interaction. 
With such a relatively weak $\Lambda\Lambda$ interaction, and since 
the three-body system $\Lambda \Lambda N$ is unbound (comparing it 
with the unbound $\Lambda nn$ system \cite{THe65}), the question of 
whether or not the onset of binding in the $S$=$-2$ hadronic 
sector occurs at $A$=$4$ becomes highly topical. 

The Brookhaven AGS experiment E906, studying $\Xi^-$ capture following 
the $(K^-,K^+)$ reaction on $^9$Be, has recently given evidence for 
excess pions that defied known single-$\Lambda$ hypernuclear weak decays 
and were conjectured as due to the formation of 
$_{\Lambda\Lambda}^{~~4}$H ($I$=0,$J^\pi$=1$^+$) \cite{Ahn01}. 
A subsequent theoretical study \cite{KFO02} 
of the weak-decay modes available to $_{\Lambda\Lambda}^{~~4}$H 
does not support this conjecture and in our opinion the question of 
whether or not $_{\Lambda\Lambda}^{~~4}$H is particle stable remains 
experimentally open. If it is confirmed in a future extension of 
E906 or of a related experiment, then this four-body system 
$\Lambda\Lambda pn$ would play as a fundamental role for studying 
theoretically the hyperon-hyperon forces as the $^{3}_{\Lambda}$H 
bound state of the three-body system $\Lambda pn$ has played 
for studying theoretically the hyperon-nucleon forces (Ref. \cite{NKG02} 
and references cited therein). 
Our aim in this Letter is to search 
theoretically for a possible $_{\Lambda\Lambda}^{~~4}$H bound state 
by solving the appropriate Faddeev-Yakubovsky equations for the 
four-body system $\Lambda\Lambda pn$, particularly for $\Lambda\Lambda$ 
interactions which reproduce the recently deduced binding energy of 
$_{\Lambda\Lambda}^{~~6}$He \cite{FGa02}. This is the first ever 
systematic Faddeev-Yakubovsky calculation done for the $A$=$4$, 
$S$=$-2$ problem. It has the virtue of taking into account properly 
{\it all} the rearrangement channels (or equivalently clusters) which 
the $\Lambda \Lambda pn$ system may be split into. 
We note that a $^{3}_{\Lambda}$H bound 
state does not necessarily imply, for attractive $\Lambda \Lambda$ 
interactions, that $_{\Lambda\Lambda}^{~~4}$H is particle stable. 

The $\Lambda N$ and $\Lambda\Lambda$ interaction potentials used 
as input were of 
a three-range Gaussian $s$-wave form similar to that used by Hiyama 
{\it et al.} \cite{HKM97,HKM02}: 
\begin{equation}
V^{(2S+1)}(r)=\sum_i^3v_i^{(2S+1)}\exp(-\frac{r^2}{\beta_i^2})\;\;. 
\label{eq:OBE}
\end{equation}
The values of the range parameters $\beta_i$ and of the 
singlet- and triplet- strength parameters $v_i^{(2S+1)}$ 
are listed in Table \ref{1}. The $\Lambda\Lambda$ interaction, 
respecting the Pauli principle, is limited to the singlet 
$s$-wave channel. The short-range term ($i$=$3$) provides 
for a strong soft-core repulsion and the long-range term ($i$=$1$) 
for attraction. The parameter $\gamma$, which controls the strength 
of the mid-range attractive term ($i$=$2$), was chosen such that the 
potential (\ref{eq:OBE}) reproduces the scattering length and the effective 
range for a given model as close as possible. Its appropriate values for 
$\Lambda N$ are listed in Table \ref{2} for two versions of model NSC97 
\cite{RSY99} considered realistic ones. For $\Lambda\Lambda$ we listed 
the value which was determined in Ref. \cite{FGa02} to reproduce the 
recently reported $B_{\Lambda\Lambda}(_{\Lambda\Lambda}^{~~6}\rm He)$ 
\cite{Tak01}. Also listed are the values of the scattering lengths 
for these $\Lambda N$ and $\Lambda\Lambda$ model interactions which 
obviously satisfy Eq. (\ref{eq:V1}). 
For the $pn$ triplet interaction we multiplied the $\Lambda N$ 
potential (\ref{eq:OBE}) by a factor $\alpha = 2.0685$, using 
$\gamma_{pn}^{(3)} = 1.0498$, in order to reproduce the $NN$ 
low-energy scattering parameters in this channel plus the binding 
energy of the deuteron. We used, for comparison, also the 
Malfliet-Tjon potential MT-III \cite{MTj69}. Our results are 
insensitive to which form is used.  

\begin{table} 
\caption{Range ($\beta$) and strength ($v$) parameters of the 
three-range Gaussian potential (\ref{eq:OBE}).} 
\label{1} 
\begin{tabular}{cccc}
\hline \hline 
$i$~ &~$\beta_i$ (fm)~&~$v_i^{(1)}$ (MeV)~&~$v_i^{(3)}$ (MeV)~ \\ 
\hline
 1~&~1.342~&~$-21.49$~&~$-21.39$~ \\ 
 2~&~0.777~&~$-379.1 \times \gamma^{(1)}$~&~$-379.1 \times \gamma^{(3)}$~ \\ 
 3~&~0.350~&~9324~&~11359~ \\
\hline \hline 
\end{tabular} 
\end{table} 

\begin{table} 
\caption{Values of the parameter $\gamma^{(2S+1)}$ appropriate for simulating 
the $\Lambda p$ potentials of model NSC97 and for a $\Lambda\Lambda$ 
potential reproducing $B_{\Lambda\Lambda}(_{\Lambda\Lambda}^{~~6}$He). 
The resulting scattering lengths $a$ (in fm) are also listed.} 
\label{2} 
\begin{tabular}{ccccc} 
\hline \hline 
Model~&~$\gamma^{(1)}$~&~$^{1}a$~&~$\gamma^{(3)}$~&~$^{3}a$~ \\
\hline 
$\Lambda N$: NSC97e~&~1.0133~&~$-2.10$~&~1.0629~&~$-1.84$~  \\ 
$\Lambda N$: NSC97f~&~1.0581~&~$-2.50$~&~1.0499~&~$-1.75$~  \\ 
$\Lambda\Lambda$: $_{\Lambda\Lambda}^{~~6}$He \cite{FGa02}~
&~0.6598~&~$-0.77$~&~--~&~--~  \\
\hline \hline 
\end{tabular}  
\end{table} 

\section{Faddeev-Yakubovsky calculations}

\begin{figure}[t] 
\centerline{\includegraphics[height=6.8cm]{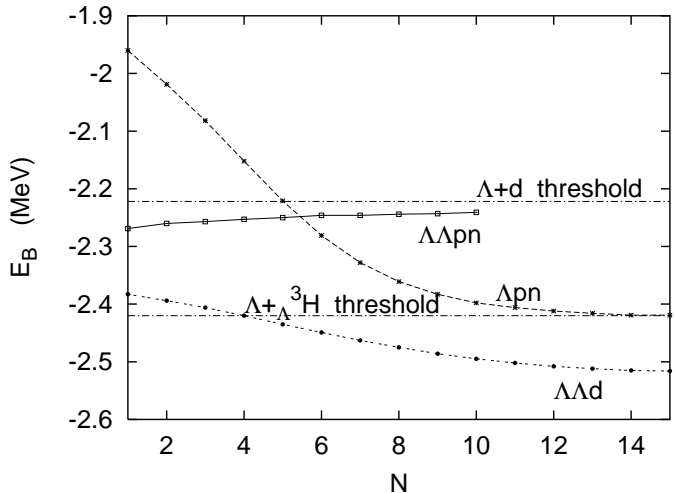}} 
\caption{Convergence of Faddeev-Yakubovsky calculations for the 
binding energy of the $\Lambda pn$ ($S$=1/2), $\Lambda\Lambda d$ 
and $\Lambda\Lambda pn$ ($S$=1) systems with respect to the number 
$N$ of basis functions. Values of $R_{\rm cutoff}=30$ fm for 
$\Lambda\Lambda pn$ and $\Lambda pn$, and $60$ fm for $\Lambda\Lambda d$ 
were used. The $\Lambda\Lambda$ interaction is due to Table \ref{2}.} 
\label{fig:BE} 
\end{figure}

We solved the differential Faddeev equations under the $s$-wave 
approximation \cite{FYa00a} for the $I$=$0$, $J^{\pi}$=$\frac12^+,\frac32^+$ 
ground-state doublet levels of $_{\Lambda}^3$H viewed as a $\Lambda pn$ system. 
Similar calculations for three-body systems are discussed in Ref. \cite{FGa02}. 
Some of our results are displayed in Table \ref{3}. The $\frac12^+$ ground 
state is bound and the calculated binding energies of the $\Lambda$ hyperon 
($B_{\Lambda}$) are in rough agreement with that observed. For model NSC97f, 
for example, our calculated $B_{\Lambda}=0.19$ MeV 
agrees with that of the recent Hiyama {\it et al.} \cite{HKM02a} 
where no $s$-wave approximation was invoked. The impact of the higher 
partial waves for $_{\Lambda}^3$H was estimated by Cobis {\it et al.} 
\cite{CJF97} to be of order 0.02 MeV, well within the error of the 
measured binding energy. Our $B_{\Lambda}$ values satisfy the 
effective-range expansion in terms of $\Lambda d$ low-energy 
parameters which are close to those derived using 
Effective Field Theory methods \cite{Ham02}. The convergence of the Faddeev 
calculation using model NSC97f for the $\Lambda N$ interaction is exhibited 
in Fig. \ref{fig:BE} as function of the number $N$ of basis functions. The 
corresponding curve, marked `$\Lambda pn$', gives the $B_{\Lambda}$ value 
with respect to the horizontal straight line marked `$\Lambda + d$ threshold'. 
The `$\Lambda pn$' asymptote serves then for defining the lowest 
particle-stability threshold, that of $\Lambda + _{\Lambda}^3$H, in the 
four-body $\Lambda\Lambda pn$ calculation described below. The $\frac32^+$ 
(unobserved and probably unbound) excited state of $_{\Lambda}^3$H comes out 
very weakly bound in our Faddeev calculation in both versions $e$ and $f$ of 
model NSC97. In order to check the sensitivity of the four-body calculation 
to the location of $_{\Lambda}^3$H($\frac32^+$) we will give below results 
also for model NSC97f', where $f'$ coincides with $f$ for the $\frac12^+$ 
channel but slightly departs from it for the $\frac32^+$ channel as shown 
in Table \ref{3}. 

\begin{table} 
\caption{$B_{\Lambda}(_\Lambda^3{\rm H}(\frac12^+))$ and $\Lambda d$ 
low-energy doublet scattering parameters ($^2B_\Lambda$ in MeV; 
$^2a$, $^2r$ in fm) calculated for the $I=0$ ${\Lambda pn}$ system. 
$^4B^{\rm scatt}_\Lambda$ (in MeV) for $_\Lambda^3{\rm H}(\frac32^+)$ 
was obtained using the effective-range 
expansion in the quartet $\Lambda d$ channel.} 
\label{3} 
\begin{tabular}{ccccc} 
\hline \hline 
 Model~ &~ $^2a$~ &~ $^2r$~ &~ $^2B_{\Lambda}$~ &~ 
$^4B^{\rm scatt}_{\Lambda}$~ \\ \hline 
 NSC97e~ &~ 20.7~ &~ 2.61~ &~ 0.069~ &~ 0.015~ \\ 
 NSC97f~ &~ 13.1~ &~ 2.46~ &~ 0.193~ &~ 0.003~ \\                 
 NSC97f'~&~ 13.1~ &~ 2.46~ &~ 0.193~ &$-0.003$~ \\ 
  ~~      &  ~~    & ~~     & ~~      &  ~~     \\ 
 EFT \cite{Ham02}&~ 16.8$^{+4.4}_{-2.4}$~ &~ 2.3$\pm$0.3~ 
&~ 0.13$\pm$0.05~ &~ \\ 
 exp. ~    & ~  & ~    &~ 0.13$\pm$0.05~ &~ \\ 
\hline \hline 
\end{tabular} 
\end{table} 

Focusing on the $\Lambda\Lambda pn$ Faddeev-Yakubovsky calculation we note 
that for two identical hyperons and two essentially identical nucleons 
(upon introducing isospin), as appropriate to the $I$=$0$, $J^{\pi}$=$1^+$ 
ground state of $^{~~4}_{\Lambda\Lambda}$H, the 18 Faddeev-Yakubovsky 
components which satisfy coupled equations reduce to seven 
independent components, in close analogy to the Faddeev-Yakubovsky 
equations discussed in our recent work \cite{FGa02} for the 
$\Lambda\Lambda\alpha\alpha$ model of $^{~10}_{\Lambda\Lambda}$Be. 
Six rearrangement channels are involved in our $s$-wave calculation 
for $^{~~4}_{\Lambda\Lambda}$H: 
\begin{equation} 
(\Lambda NN)_{S=\frac12} + \Lambda \;, \;\; 
(\Lambda NN)_{S=\frac32} + \Lambda \;, \;\; 
(\Lambda \Lambda N)_{S=\frac12} + N \; 
\label{eq:3+1} 
\end{equation}
for 3+1 breakup clusters, and 
\begin{equation} 
(\Lambda \Lambda)_{S=0} + (NN)_{S=1} \;, \;\;\; 
(\Lambda N)_S + (\Lambda N)_{S'} 
\label{eq:2+2} 
\end{equation} 
with $(S,S')$=$(0,1)$+$(1,0)$ and $(1,1)$ for 2+2 breakup clusters.  
We find invariably that the three rearrangement channels in which 
the two nucleons belong to the same $d$-like cluster dominate in 
actual calculations. This observation, apparently, could justify 
the use of a $\Lambda\Lambda d$ model for $^{~~4}_{\Lambda\Lambda}$H. 
However, as we shall see and discuss below, the results of using such 
a three-body model differ radically from those of the full four-body 
Faddeev-Yakubovsky calculations which retain 
the proton and neutron as dynamically independent entities. 

\section{Results}

Using the $\Lambda\Lambda$ interaction which reproduces 
$B_{\Lambda\Lambda}(^{~~6}_{\Lambda\Lambda}$He) (see Table \ref{2}) 
our calculations yield no bound state for the $\Lambda\Lambda pn$ 
system, as demonstrated in Fig. \ref{fig:BE} by the location of the 
`$\Lambda\Lambda pn$' curve {\it above} the horizontal straight line 
marking the `$\Lambda + ~_{\Lambda}^{3}{\rm H}$ threshold' \cite{com02}. 
In fact our Faddeev-Yakubovsky calculations exhibit little sensitivity 
to the strength of the $\Lambda\Lambda$  interaction over a wide range, 
including much stronger $\Lambda\Lambda$ interactions such as ND and ESC00 
\cite{FGa02}, the latter one reproducing the (excessive) $B_{\Lambda\Lambda}$ 
value reported for the `old' $^{~~6}_{\Lambda\Lambda}$He event \cite{Pro66}. 
For these $\Lambda\Lambda$ interactions we get a bound 
$^{~~4}_{\Lambda\Lambda}$H only if the $\Lambda N$ interaction is made 
considerably stronger, by as much as 40\%. With four $\Lambda N$ pairwise 
interactions out of a total of six, the strength of 
the $\Lambda N$ interaction (here about half of that for $NN$ binding 
the deuteron) plays a major role in the four-body $\Lambda\Lambda pn$ 
problem. In passing we remark that this is also apparent from the bounds 
derived in Ref. \cite{JMR94} for the four-body bound-state problem. Put 
differently, we know of no few-body theorem 
that would imply, for essentially attractive $\Lambda\Lambda$ interactions 
and for a {\it non static} nuclear core $d$ (made out of $pn$ in the present 
case), the existence of a $\Lambda\Lambda d$ bound state provided that 
$\Lambda d$ is bound. It is a remarkable outcome of the complete 
Faddeev-Yakubovsky scheme for four particles that such a natural expectation 
can be refuted by a specific calculation. However, for a {\it static} 
nuclear core $d$, and disregarding inessential complications due to spin, 
a two-body $\Lambda d$ bound state does imply binding for the three-body 
$\Lambda\Lambda d$ system \cite{Bas02}. A discussion of the formal 
relationship between these four-body and three-body models which do not 
share a common hamiltonian is deferred to a subsequent publication. 

Our $\Lambda\Lambda d$ model for $^{~~4}_{\Lambda\Lambda}$H uses the 
$\Lambda\Lambda$ interaction marked `$^{~~6}_{\Lambda\Lambda}$He' 
in Table \ref{2} plus $\Lambda d$ interactions that reproduce the 
low-energy parameters of the $\Lambda pn$ Faddeev calculation specified 
in Table \ref{3}. The dependence on the functional form chosen for the 
interpolating $\Lambda d$ interaction potentials proved relatively mild. 
The results of such a $\Lambda\Lambda d$ three-body 
Faddeev calculation using model NSC97f for the underlying $\Lambda N$ 
interaction are shown in Fig. \ref{fig:BE} as function of the number $N$ 
of basis functions used in the expansion of the Faddeev components. 
The asymptote of the curve marked `$\Lambda\Lambda d$' is now located 
{\it below} the horizontal straight line for the 
`$\Lambda + ~_{\Lambda}^{3}{\rm H}$ threshold', 
so $^{~~4}_{\Lambda\Lambda}$H is particle stable. 
The figure may suggest that a $\Lambda$ in $^{~~4}_{\Lambda\Lambda}$H 
is less bound, by about 0.1 MeV, than a $\Lambda$ in $_{\Lambda}^{3}{\rm H}$ 
( which in model NSC97f is bound by about 0.2 MeV). However, 
the $(2J+1)$ spin-averaged effective 
$B_{\Lambda}(_{\Lambda}^{3}{\rm H})$ in $^{~~4}_{\Lambda\Lambda}$H is only 
${\bar B}_{\Lambda} = 0.07$ MeV and, since $B_{\Lambda\Lambda} \sim 0.3$ 
MeV, we have $B_{\Lambda\Lambda} > 2{\bar B}_{\Lambda}$ which is 
equivalent to stating loosely that the second $\Lambda$ in 
$^{~~4}_{\Lambda\Lambda}$H is bound even more 
strongly than the first one. This holds also for model NSC97e and it is 
a general property of the Faddeev calculation \cite{FGa02}. 

\begin{figure}[t] 
\centerline{\includegraphics[height=6.8cm]{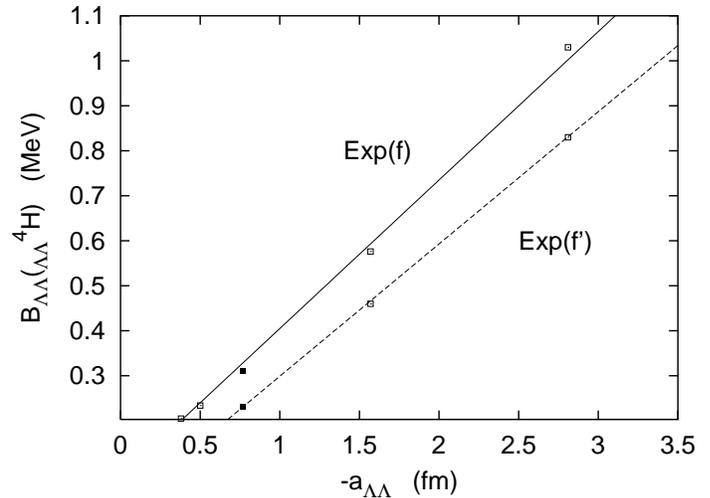}} 
\caption{$B_{\Lambda\Lambda}(_{\Lambda\Lambda}^{~~4}\rm H)$ calculated 
in a three-body $\Lambda\Lambda d$ model as function of the scattering 
length $a_{\Lambda\Lambda}$, for two exponential $\Lambda d$ 
potentials corresponding to versions $f$ and $f'$ in Table \ref{3}. 
The solid squares correspond to the 
`$^{~~6}_{\Lambda\Lambda}$He' $\Lambda\Lambda$ interaction of Table \ref{2}. 
The straight lines are drawn only to lead the eye.} 
\label{fig:Expf} 
\end{figure} 

In Fig. \ref{fig:Expf} we show $B_{\Lambda\Lambda}$ values 
calculated for $^{~~4}_{\Lambda\Lambda}$H within this $\Lambda\Lambda d$ 
Faddeev model as function of ${\bar V}_{\Lambda\Lambda}$ 
(quantified by the value of the scattering length $a_{\Lambda\Lambda}$) 
for two exponential $\Lambda d$ potentials corresponding 
to versions $f$ and $f'$ of model NSC97 (see Table \ref{3}). 
The roughly linear increase of $B_{\Lambda\Lambda}$ holds generally 
in three-body $\Lambda \Lambda C$ models ($C$ standing for a cluster) 
over a wide range of values for ${\bar V}_{\Lambda\Lambda}$ \cite{FGa02}. 
For values $B_{\Lambda\Lambda}\leq 0.2$ MeV, 
$^{~~4}_{\Lambda\Lambda}$H becomes unstable against emitting a $\Lambda$. 
This onset of particle stability for $^{~~4}_{\Lambda\Lambda}$H requires 
a minimum strength for the $\Lambda\Lambda$ interaction which is satisfied 
for our choice of $^{~~6}_{\Lambda\Lambda}$He as a normalizing datum. It is 
also seen from the figure that the uncertainty in the location of 
$^3_{\Lambda}{\rm H}(\frac32^+)$ bears serious consequences for the 
predicted binding of $^{~~4}_{\Lambda\Lambda}$H; this is a particularly 
strong effect as the $\frac32^+$ state crosses the $\Lambda + d$ 
threshold. Yet, we would like to emphasize that no such sensitivity emerges 
within a genuine four-body model calculation which does not bind 
$^{~~4}_{\Lambda\Lambda}$H as long as the $\Lambda N$ interaction is of the 
size constrained by single-$\Lambda$ hypernuclear phenomenology. 

\section{Discussion and conclusions}

In cluster models of the type $\Lambda\Lambda C$ and $\Lambda\Lambda C_1 C_2$ 
for heavier $\Lambda\Lambda$ hypernuclei, where the nuclear-core cluster 
$C$=$C_1$+$C_2$ is made out of subclusters $C_1$ and $C_2$, the $\Lambda C_j$ 
interaction (normally producing bound states) is considerably stronger than 
for $\Lambda N$. 
Our experience with Faddeev-Yakubovsky calculations for 
$^{~10}_{\Lambda\Lambda}$Be \cite{FGa02}, viewed as a four-body 
$\Lambda\Lambda\alpha\alpha$ system, 
is that the relationship between the three-body and four-body models 
is then opposite to that found here for $^{~~4}_{\Lambda\Lambda}$H: 
the $\Lambda\Lambda C_1 C_2$ calculation under 
similar conditions provides {\it higher} binding than the $\Lambda\Lambda C$ 
calculation yields. The mechanism behind it is the attraction induced by 
the $\Lambda C_1$-$\Lambda C_2$, $\Lambda\Lambda C_1$-$C_2$, 
$C_1$-$\Lambda\Lambda C_2$ four-body rearrangement channels which include 
bound states that have no room for in the three-body $\Lambda\Lambda C$ model. 
The binding energy calculated within the four-body model increases then 
`normally' with ${\bar V}_{\Lambda\Lambda}$. 

In conclusion, we have provided a first four-body Faddeev-Yakubovsky 
calculation for $^{~~4}_{\Lambda\Lambda}$H using $NN$ and $\Lambda N$ 
interaction potentials that fit the available data on the relevant 
subsystems, including the binding energy of $^3_\Lambda$H. 
No bound state is obtained for $^{~~4}_{\Lambda\Lambda}$H 
over a wide range of $\Lambda\Lambda$ interaction strengths, 
including that normalized to reproduce the binding energy of 
$^{~~6}_{\Lambda\Lambda}$He. We have traced the origin of 
this non binding as due to the relatively weak $\Lambda N$ 
interaction. This is in stark contrast to the results of 
a `reasonable' three-body $\Lambda\Lambda d$ Faddeev 
calculation that binds $^{~~4}_{\Lambda\Lambda}$H provided 
the $\Lambda\Lambda$ interaction is not too weak, 
say with $-a_{\Lambda\Lambda} \geq 0.5$ fm.

\begin{acknowledgments} 
A.G. gratefully acknowledges useful correspondence with Toshio 
Motoba and with Jean-Marc Richard. 
This work was partially supported by the Israel Science Foundation. 
I.N.F. was also partly supported by the Russian Foundation for Basic 
Research (grant No. 02-02-16562). 
\end{acknowledgments}

\end{document}